\documentclass[%
  reprint,
superscriptaddress,
preprintnumbers,
bibnotes,
 aps,
 prl,%
]{revtex4-2}

\usepackage{graphicx}
\usepackage{dcolumn}
\usepackage{bm}
\usepackage[mathlines]{lineno}

\usepackage{xcolor,soul}
\sethlcolor{white}

\usepackage[version=3]{mhchem}

\makeatletter
\newcommand{\customlabel}[2]{%
\protected@write \@auxout {}{\string \newlabel {#1}{{#2}{}}}}
\makeatother

\begin{document}


\title{
Switchable Electric Dipole from Polaron Localization in Dielectric Crystals
}

\date{\today}

\author{Kazuki Morita}
\affiliation{Department of Materials, Imperial College London, London SW7 2AZ, United Kingdom }

\author{Yu Kumagai}
\affiliation{Laboratory for Materials and Structures, Institute of Innovative Research, Tokyo Institute of Technology, Yokohama 226-8503, Japan}

\author{Fumiyasu Oba}
\affiliation{Laboratory for Materials and Structures, Institute of Innovative Research, Tokyo Institute of Technology, Yokohama 226-8503, Japan}
\affiliation{Materials Research Center for Element Strategy, Tokyo Institute of Technology, Yokohama 226-8503, Japan}

\author{Aron Walsh}
\email{a.walsh@imperial.ac.uk}
\affiliation{Department of Materials, Imperial College London, London SW7 2AZ, United Kingdom }
\affiliation{Department of Materials Science and Engineering, Yonsei University, Seoul 03722, Korea}

\begin{abstract} 
Ferroelectricity in crystals is associated with the displacement of ions or rotations of polar units. Here we consider the dipole created by donor doping ($D^+$) and the corresponding bound polaron ($e^-$). A dipole of 6.15 Debye is predicted, from Berry phase analysis, in the Ruddlesden-Popper phase of \ce{Sr3Ti2O7}. A characteristic double-well potential is formed, which persists for high doping densities. The effective Hubbard $U$ interaction can vary the defect state from metallic, a two-dimensional polaron, through to a zero-dimensional polaron. The ferroelectric-like behavior reported here is localized and distinct from conventional spontaneous lattice polarization.


\end{abstract}


\maketitle

The ability to switch polarization by an external electric field makes ferroelectric crystals an essential technological system \cite{LinesGlass}.
One difficulty in developing new materials is the complexity of contributions coupled over a range of length scales \cite{Jia2006}.
The prominent example is where sample shape and atomic displacements interplay and cause ferroelectricity to diminish with decreasing film thickness, typically seen in perovskites \cite{Fong2004}.
The field has evolved with the development of hybrid improper ferroelectrics and multi-ferroelectrics \cite{Benedek2011,Harris2011,Bousquet2008,Garrity2014,Martin2016}.
These non-conventional cases exhibit different interactions with other physical quantities, notably the strain field, and are thus anticipated to overcome existing performance bottlenecks \cite{Cheema2020,Spreitzer2021}.

Beyond stoichiometric materials, recent efforts have focused on the interplay of defects with ferroelectricity \cite{Ren2004}.
Despite the rudimentary electrostatic theory suggesting that the free electrons or holes induced by defects screen the internal polarization and suppress ferroelectricity, there are cases where defects can facilitate ferroelectricity \cite{Li2021,Kolodiazhnyi2010}.
One example is stabilization of the ferroelectric phase of \ce{HfO2} by doping \cite{Muller2011,Mueller2012,Hoffmann2015}.
Several studies have also suggested the possibility of inducing polar distortions through doping \cite{Ricca2021,Li2021}.
For example, Ricca et al. have studied complex of oxygen and strontium vacancy in \ce{SrMnO3} and reported that they are capable of creating a switchable dipole \cite{Ricca2021}.
Li and Birol have shown that electron doping could stabilize hybrid-improper octahedral rotation in Ruddlesden-Popper phase compounds \cite{Li2021}.
Efforts have also been made to enhance dielectric screening using defects.
Although the polarization does not persist, colossal permittivity has attracted considerable attention \cite{Hu2013,Hu2015,Dong2015,Berardan2016}.
Microscopically, this behavior is realized by the complex of positively and negatively charged defects.
Dipoles can reorient when the defects rearrange their configuration upon radiation of an external electric field \cite{Hu2013}.
However, most colossal dielectric materials require three or four types of defects to be located in close proximity, therefore, finding simpler alternatives is desirable.

In this Letter, we predict the activation of ferroelectric-like behavior in a non-polar host crystal.
This is achieved through an F donor (+ charge) and the corresponding electron polaron ($-$ charge) in \ce{Sr3Ti2O7}.
Using Berry phase analysis and density functional theory (DFT), we show that the F-polaron complex could induce a finite switchable dipole.
To understand the diversity of Ruddlesden-Popper phases \cite{Mulder2013}, we model possible accessible behavior through tuning of the Hubbard $U$ parameter in \hl{the} exchange-correlation functional.
Although there are many similarities to conventional ferroelectrics, the dipole in this work is based on the polaronic state.
Therefore, we expect it to have qualitatively different behavior, which may surmount existing technological bottlenecks.

\textit{Methodology:}
The modern theory of polarization requires the use of wave functions to calculate polarization \cite{Resta1992,King-Smith1993}.
The electronic contribution is defined relative to the reference state up to a polarization quanta.
Applying similar formalism towards defective systems has two difficulties.
Firstly, defects often cause partial occupancy of electronic bands, which makes polarization ill-defined.
Secondly, it is difficult to define a reference state.
In the bulk, reference structures are often taken as a midpoint between opposite polarization, which may correspond to the parent space group of the polar phase.
For the reference structures to be meaningful, the polar structures must be connected with an accessible energy barrier, otherwise the dipole cannot flip.
Given these difficulties, many studies assume a nominal point charge in place of atoms to treat polarization.
Despite the simplicity of this classical approach, errors arise from ignoring microscopic electronic contributions.
In our case, a suitable reference state can be defined.

Plane-wave DFT calculations within projector-augmented wave scheme were performed using {\it VASP} \cite{Blochl1994,Kresse1996_PRB15,Kresse1996_PRB11169}.
Using the conventional cell structure for \ce{Sr3Ti2O7}, the cell volume and the atomic coordinates were fully relaxed using the HSE06 functional \cite{Heyd2003}.
For the defect calculations, $4 \times 4 \times 1$ supercells were calculated with $1 \times 1 \times 1$ reciprocal space sampling, and cut-off energy of 450 eV was employed.
The barrier for the dipole switching was calculated with nudged elastic band, where 9 and 12 intermediate images were used for PBE+$U$ with $U$=4.0 and 5.0 eV, respectively \footnote{
The calculations of the defect formation energies, chemical potential phase diagrams and electrostatic convergence checks were done using the {\it pydefect} package \cite{Kumagai2021}.
The nudged elastic band calculations were done with {\it VASP} modified with {\it VTST} \cite{Sheppard2008}.
The Hubbard $U$ parameter was applied through the rotational invariant form introduced by Dudarev et al.\cite{Dudarev1998}
}.

\begin{figure}[tb]
\includegraphics[width=0.8\columnwidth]{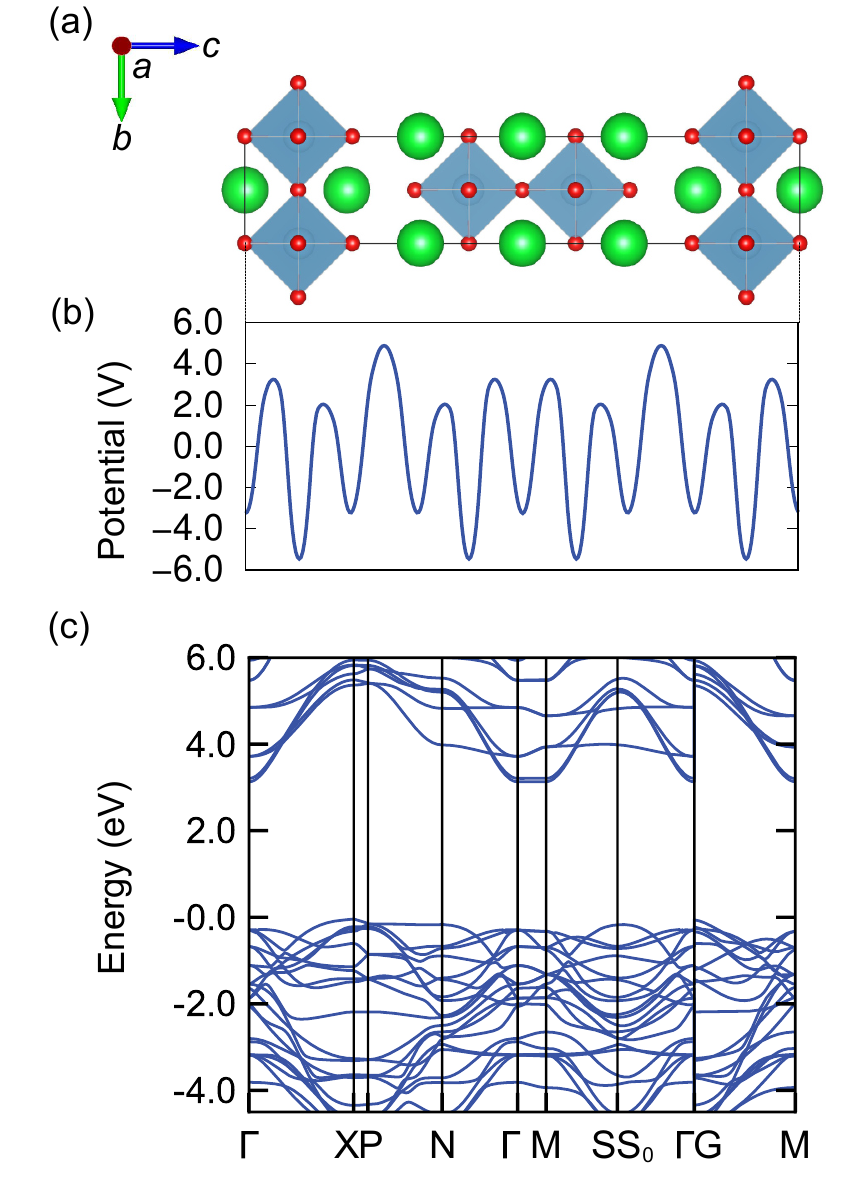}
\caption{\label{fig:struct}
  (a) Ruddlesden-Popper phase of \ce{Sr3Ti2O7}.
  The green and red circles represent strontium and oxygen atoms, respectively.
  The \ce{TiO6} octahedra are shaded in blue.
  The image was prepared with \emph{VESTA}.\cite{Momma2011}
  (b) Local Hartree potential integrated along the in-plane direction.
  (c) Electronic band dispersion. The energy zero corresponds to the valence band maximum.
  }
\end{figure}

\textit{Defect energetics:}
The structure of \ce{Sr3Ti2O7} is shown in Fig.~\ref{fig:struct}, which has a non-polar $I$4/$mmm$ space group.
This polymorph is second smallest ($n$=2) amongst Ruddlesden-Popper phases \ce{Sr_{$n$+1}Ti_$n$O_{3$n$+1}}, and has stacking of two \ce{SrTiO3} perovskite-like layers and a \ce{SrO} rocksalt-like layer along the $<001>$.
The electrostatic potential for the rocksalt layer is higher than the perovskite layer (Fig.~\ref{fig:struct}(b)), which has been reported to act as an insulating layer in quantum confinement \cite{ReyesLillo2016,Li2019}.
This is mirrored in the band structure, where the conduction band dispersion is larger between the in-plane direction, $\Gamma$ and X, but smaller in the out-of-plane direction, $\Gamma$ to M (Fig.~\ref{fig:struct}(c)).

To obtain defect energies, we calculated the phase diagram with respect to the chemical potentials (see Figs.~\ref{figSI:chempotHSE06} and \ref{figSI:chempot}).
The formation energy of F calculated is 2.19 eV (HSE06), whereas it was 2.59 and 2.63 for $U$=4.0 eV and $U$=5.0 eV, respectively.
Since all systems considered in this work are neutral, no charge corrections are required.
The electrostatic potential far from the defect in $4 \times 4 \times 1$ cell was well converged (Fig.~\ref{figSI:HSE06_elec}).

\begin{figure}[tb]
\includegraphics[width=\columnwidth]{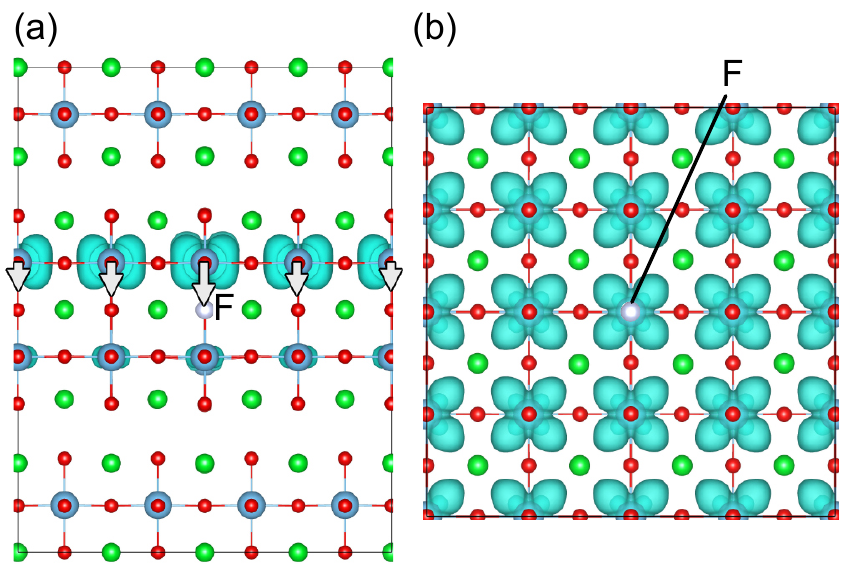}
\caption{\label{fig:polaron} 
Electron polaron density due to \ce{F_O} doping of \ce{Sr3Ti2O7} calculated by DFT/HSE06 functional viewed from (a) an in-plane direction, and (b) an out-of-plane direction.
Green, blue, red, and gray circles are strontium, titanium, oxygen, and fluorine, respectively.
Gray arrows are a guide to the eye for the dipole direction.
  }
\end{figure}

\textit{Polaron distribution:}
As the pristine system is diamagnetic, the spin density is a useful descriptor of the unpaired electron in the polaronic state.
The HSE06 analysis is shown in Figs.~\ref{fig:polaron}(a) and (b) (results for PBE+$U$ are shown in Figs.~\ref{figSI:PBEU40_pol} and \ref{figSI:PBEU50_pol}).
The polaron exhibited two-dimensional (2D) localization where it was well localized in the out-of-plane direction (Fig.~\ref{fig:polaron}(a)), but was spread widely along the in-plane direction (Fig.~\ref{fig:polaron}(b)).
This behavior is similar to the 2D excitons described by the Bethe-Salpeter equation \cite{ReyesLillo2016}.
The anisotropic dielectric screening also explains this behavior (Table~\ref{tabSI:dielectric_tensor}).
The 2D polaron had a slightly higher density in the proximity of the F donor, suggesting a finite radius.
However, even in our $6 \times 6 \times 1$ supercell, it was not possible to fully encompass the spread of the 2D polaron (Fig.~\ref{figSI:PBEU40_pol}).

We can estimate the diameter by using the Fr\"ohlich polaron model for isotropic media.
Following Schultz, the polaron radius $r_f$ was calculated as \cite{Schultz1959}:
\begin{equation}
r_f = \frac{3 v^2}{2 v m (v^2 - w^2)},
\end{equation}
where $v$ and $w$ are calculated solely using Feynman's theory \cite{Feynman1955,Frost2017}, $m=0.12$ is the electron effective mass.
The coupling strength $\alpha$ is defined as:
\begin{equation}
\alpha = \frac{1}{2}\left( \frac{1}{\varepsilon_\infty} - \frac{1}{\varepsilon_0} \right) \frac{e^2}{\hbar \omega_{\rm LO}} \left( \frac{2m\omega_{\rm LO}}{\hbar} \right)^{1/2},
\end{equation}
where $\varepsilon_\infty=30.61$ is the high-frequency dielectric constant, $\varepsilon_0=5.25$ is the low-frequency dielectric constant, $e$ is the elementary charge, and $\omega_{\rm LO}$ is the longitudinal optical phonon frequency.
The resulting $\alpha$ was 3.81 and was used to calculate $v$ and $w$ (detail shown in the Supplementary Material).
The longitudinal optical phonon modes at the zone-center ($\Gamma$-point) was calculated using \emph{phonopy} \cite{Togo2015,Skelton2017}, and they were averaged using the Hellwarth and Biaggio method \cite{Hellwarth1999,Frost2017}.
The resulting polaron radius $r_f$ was 52.80 \AA, which was much larger than the size of $6\times 6 \times 1$ supercell having 23.4 \AA\  in the in-plane direction.

We also considered a continuum electrostatic model:
\begin{equation}
E(\psi) = \int d{\bm r} \left[ \psi^*({\bm r}) \left( -\frac{\hbar \nabla^2}{2m} \right)\psi({\bm r}) - \frac{1}{2}{\bm E}({\bm r})\cdot {\bm D}({\bm r}) \right]
\end{equation}
Here $E(\psi)$ is the energy, $\psi$ is the polaron wavefunction, ${\bm E}$ is the self-consistent electric field, and ${\bm D}$ is the electronic displacement field by the polaron and the medium was assumed to be isotropic in three dimensions \cite{Sio2019,Devreese2009}.
The polaron radius $r_p$ was obtained by minimizing the above energy with respect to the trial wavefunction $\psi({\bm r}) = (\pi r_p^3)^{-1/2} e^{-r/r_p}$, where $r_p$ is the polaron radius.
The obtained radius was 56.86 \AA, which was a similar value to the result using Schultz's formalism.

We then modeled the dipole induced by the complex of \ce{F^+} and the polaron.
Since formation of polaron breaks the inversion symmetry, it is tempting to use the non-relaxed structure as a reference structure, where the excess charge induced by the dopant is symmetrically distributed.
However, we found that this structure was metallic, so the polarization is ill-defined.
An artificial structure, where the bonding around Ti was expanded to localize polaron symmetrically was used.
The reference polarization vanishes with modulo of $e{\bf R}/\Omega$ \cite{Vanderbilt1993}.

The resulting dipole was 6.15 Debye (spontaneous polarization $P_S$=0.41 $\mu C/cm^2$ in $4\times4\times1$ supercell).
This value is small compared to prototypical ferroelectrics \cite{LinesGlass}, but direct comparison is not straightforward.
A better quantification could be made by considering a point charge model.
By placing a $+1$ and a $-1$ charge in the location of the F and the neighboring Ti in the unrelaxed supercell (1.92 \AA\ apart) , the resulting classical dipole is 9.52 Debye.
The reduction of 35\% compared to the DFT calculation can be attributed to dielectric screening.

\begin{figure}[tb]
\includegraphics[width=0.8\columnwidth]{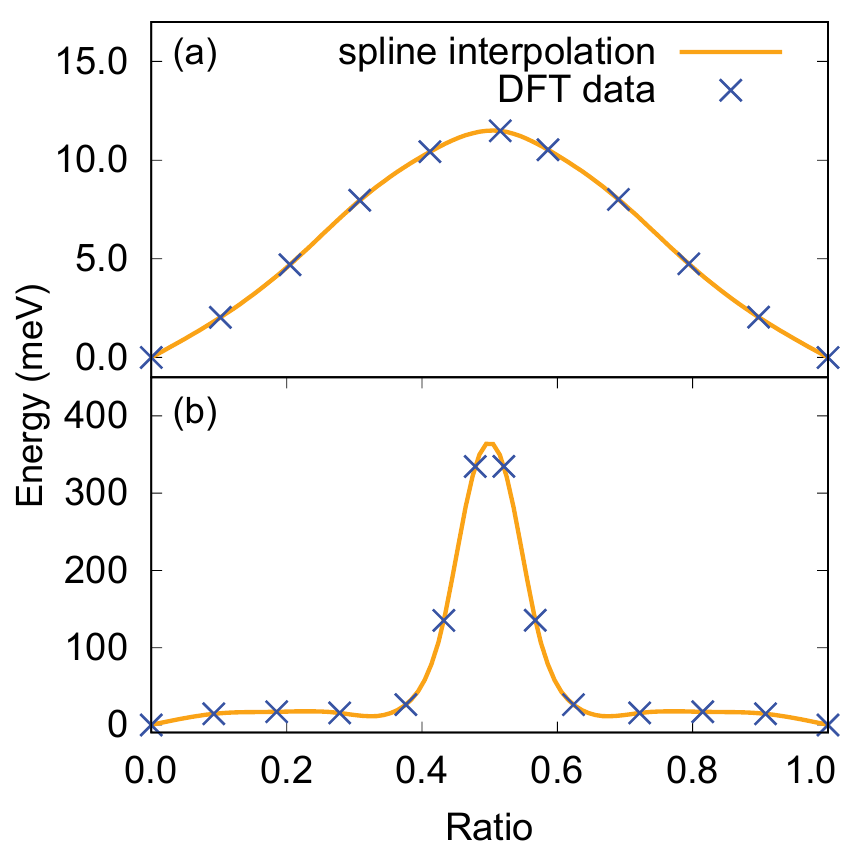}
\caption{\label{fig:neb} 
  Hopping barrier of electron polarons is calculated with (a) PBE+$U$ ($U$=4.0 eV) and (b) PBE+$U$ ($U$=5.0 eV).
  The horizontal axis was calculated by the projection along the linearly interpolated path between the initial and the final structure in the configuration space.
  }
\end{figure}

\textit{Polarization switching:}
We have shown a dipole can be formed in this system, but it must be switchable to mimic a ferroelectric response.
Based on the Landau-Devonshire model, a double-well potential should exist.
The switching barrier for the 2D polaron (Fig.~\ref{fig:neb}(a)) and zero-dimensional (0D) polaron (Fig.~\ref{fig:neb}(b)) were 11 meV and 364 meV, respectively.
A subtle change in the Hubbard $U$ parameter, $U$=4.0 to 5.0 eV, increases the hopping barrier by orders of magnitude.
This originates from the qualitative different hopping mechanisms.
As is apparent from Fig.~\ref{fig:neb}, 2D polarons gradually move to the opposite site, whereas 0D polaron stays in one site and suddenly hops.
The HSE06 calculation showed a 2D polaron structure (Fig.~\ref{fig:polaron}), so Fig.~\ref{fig:neb}(a) is closer, however as we will later discuss, 0D polarons may be accessible by composition engineering.
Longer-rang hopping through the rocksalt layer into the next nearest perovskite layer was too unfavourable to realize.
Such a barrier may not be present in the less anisotropic structure of \ce{TiO2}, where many of the colossal permittivity studies have been performed \cite{Hu2013}, and suggests that dielectric loss may be reduced.

A double-well potential may not be realized for all combinations of dopants and suitable host materials, because the binding energy of the polaron must be in an optimal range.
If bound too strongly, the hopping barrier will diminish to a single-well structure.
On the other hand, if the binding is too small, the polaron will diffuse away and cause the dipole to collapse.
The results show that polarons in \ce{Sr3Ti2O7} fall in the optimal range.


\begin{figure}[tb]
\includegraphics[width=0.9\columnwidth]{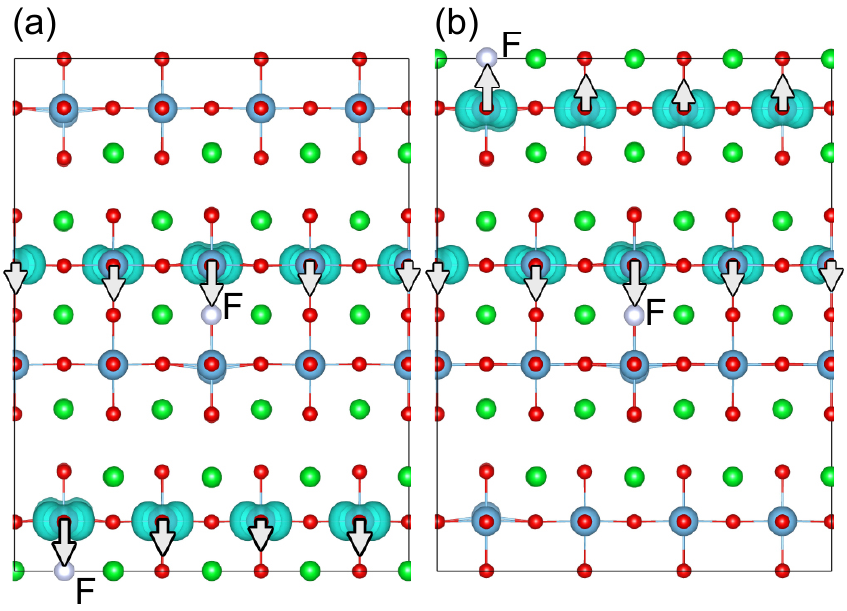}
\caption{\label{fig:double_defect} 
  Polaron distribution for the case of two \ce{F_O} in a $4 \times 4 \times 1$ supercell of \ce{Sr3Ti2O7}.
  In case of (a) ``ferroelectric'' and (b) ``antiferroelectric'' configurations.
  Green, blue, red, and gray circles are strontium, titanium, oxygen, and fluorine, respectively.
  Gray arrows are guides to the eye for the dipole directions.
  }
\end{figure}

\textit{Higher doping levels:}
It is worthwhile to the effect of
higher polaron concentrations.
Fig.~\ref{fig:double_defect} shows the result for a doubled defect density.
The ``antiferroelectric'' configuration (Fig.~\ref{fig:double_defect}(b)) is 0.6 meV more stable than the ``ferroelectric'' dipole configuration (Fig.~\ref{fig:double_defect}(a)).
If each perovskite bi-layer is considered as a domain, the energy difference could be converted to interfacial energy of 0.018 mJ/$m^2$ (0.0012 meV/\AA$^2$).
This energy is orders of magnitudes smaller than that seen for conventional ferroelectric materials, such as \ce{BaTiO3}, where the interfacial energy is in the order of $\sim$10 mJ/$m^2$ \cite{Marton2010,Grunebohm2012,Grunebohm2020}.
The small energy highlights the fundamentally different mechanism of the ferroelectricity in the F-polaron dipole system, which relies largely on the local electronic structure rather than the long-range displacement of ions.

The ``ferroelectric'' configuration has a dipole strength of 8.37 Debye, while it vanishes for the
``antiferroelectric'' configuration.
Since the single F-polaron pair had 6.15 Debye, the dipole did not double with doping density.
Although the interaction energy between the neighboring dipole was small, this result suggests that electrostatic repulsion between the polarons across the rocksalt layer is present.

\begin{figure}[tb]
\includegraphics[width=0.9\columnwidth]{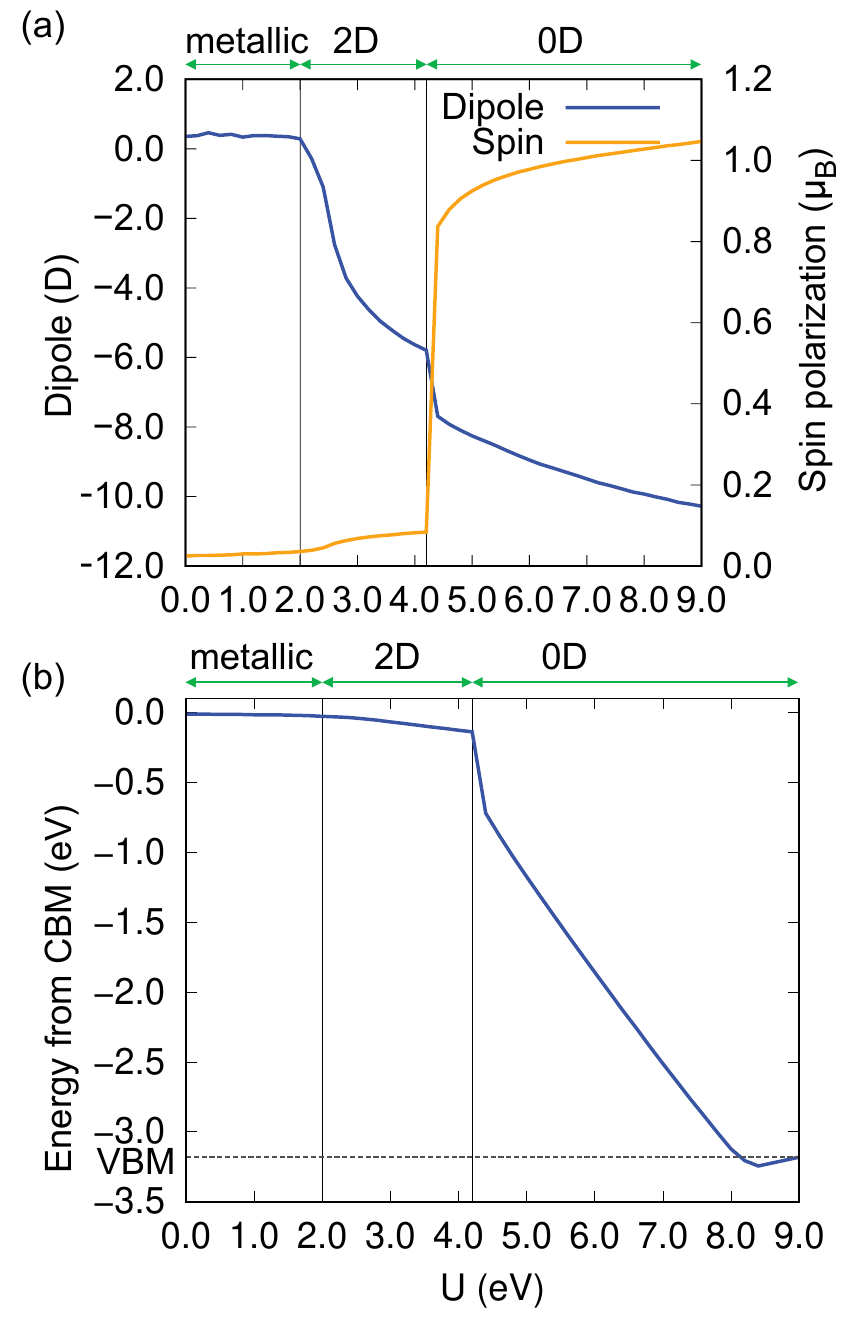}
\caption{\label{fig:U_param}
  (a) Relation between the Hubbard $U$ parameter and, the total dipole and spin polarization in the polaron location.
  (b) Defect state depth from conduction band minimum (CBM) (full density of states presented in Fig.~\ref{figSI:full_dos}).
  The labeled VBM is the value of the valence band maximum at $U$=9.0 eV.
  The label ``metallic'', ``2D'', and ``0D'' corresponds to the respective polaron solutions, and their boundaries are drawn with a vertical line at 2.0 eV and 4.2 eV.
  }
\end{figure}

\textit{Polaron regimes:}
An extended family of Ruddlesden-Popper phases exist \cite{Mulder2013}.
Instead of performing exhaustive calculations, we model the extremes of behavior by varying the Hubbard $U$ parameter.
Such a variation could be realized by changing the B-site cation; effective Hubbard $U$ values for 3d metals range from $\sim$2.5 eV in Sc to $\sim$13.0 eV in Ni \cite{Torrance1991,Imada1998,Aryasetiawan2006}.
We note that surfaces could alter the effective $U$ values through modification of the atomic environments \cite{Wehling2011}.

Three distinct segments of the curve can be discerned in Fig.~\ref{fig:U_param}(a).
The first is when the defect state falls in the conduction band and acts as an electron donor.
Here delocalization throughout the crystal is seen (Fig.~\ref{figSI:polaron_PBEU}(a)).
The system is metallic and dipoles are fully screened.
The slight deviation of polarization from 0 in Fig.~\ref{fig:U_param}(a) is due to limiations of the formalism. 
The second regime starts from about $U$=2.0 eV and continues up to $U$=4.2 eV.
This corresponds to a 2D polaron solution (Fig.~\ref{figSI:polaron_PBEU}(b)) and corresponds to polaron shape from the HSE06.
From this value of $U$, a dipole emerges as a result of broken inversion symmetry by localization of polaron on a single side of the donor.
Just over $U$=4.2 eV, the dipole strength discontinuously changes, and the polaron becomes 0D (Fig.~\ref{figSI:polaron_PBEU}(c)).
This polaron distribution is similar to the case reported in the proximity of an oxygen vacancy in \ce{SrTiO3} \cite{Janotti2014}.
Near the transition, the energy difference between 0D and 2D is small, allowing for coexistence of the two solutions.

To quantify the extent of localization, we integrated the spin density difference within the radius 1.3 \AA~sphere about each Ti.
The site with maximum magnetisation was consistently Ti atom neighboring F in positive $c$ direction and coincided with the location of the 0D polaron.
The result is overlayed in Fig.~\ref{fig:U_param}(a).
The change between the metallic occupation and 2D polaron was less apparent, which can be explained by the subtle polaron distribution change (Fig.~\ref{figSI:polaron_PBEU}(a) and (b)).
%
%

The change in the defect single-particle level is shown in Fig.~\ref{fig:U_param}(b).
As excess electrons localize in the Ti 3d, increasing the $U$ parameter has an effect of deepening the level \cite{Haldane1976}.
Again, the metallic to 2D transition is not striking, while the 2D to 0D transition is vivid.
This suggests that the former is analogous to a second-order phase transition, whereas the latter is first-order.
Over $U$=8.0 eV, the defect state reaches the valence band and becomes a resonant band.

In conclusion, we have presented the behavior of dipole created by the complex of \ce{F^+} and polaron in F-doped \ce{Sr3Ti2O7}.
The dipole behaves similarly to ferroelectrics by exhibiting double-well potential energy surfaces.
Calculation of multiple defects showed the possibility that the domain interfacial energies are orders of magnitude smaller than conventional ferroelectrics.
By tuning the Hubbard $U$ parameter, we showed three types of polaron behavior.
These results suggest the possibility of this dipole mimicking ferroelectric behavior, yet relying on a distinct microscopic mechanism.
\hl{
Additionally, the Ruddlesden-Popper phase is home to rich phenomena, including improper ferroelectricity and orbital-ordering}{\cite{Moritomo1995,Benedek2011,Martin2016}}.
\hl{
The dipole realized in this work is small compared to conventional ferroelectrics and is premature for commercial devices, but realizing a finite value from a non-polar host crystal has conceptual importance.
The influence of strain, domain effects, surface screening and choice of dopants remain to be investigated.
}

\begin{acknowledgments}
  Funding was received from the Yoshida Scholarship Foundation and Japan Student Services Organization.
  This work was also supported by the core-to-core collaboration funded by EPSRC (EP/R034540-1) and JSPS (JPJSCCA20180006).
  Via our membership of the UK's HEC Materials Chemistry Consortium, funded by EPSRC (EP/L000202 and EP/P020194), this work used the ARCHER2 Supercomputing Service.
\end{acknowledgments}

\bibliography{references}

\providecommand{\noopsort}[1]{}\providecommand{\singleletter}[1]{#1}%
\begin{thebibliography}{54}%
\makeatletter
\providecommand \@ifxundefined [1]{%
 \@ifx{#1\undefined}
}%
\providecommand \@ifnum [1]{%
 \ifnum #1\expandafter \@firstoftwo
 \else \expandafter \@secondoftwo
 \fi
}%
\providecommand \@ifx [1]{%
 \ifx #1\expandafter \@firstoftwo
 \else \expandafter \@secondoftwo
 \fi
}%
\providecommand \natexlab [1]{#1}%
\providecommand \enquote  [1]{``#1''}%
\providecommand \bibnamefont  [1]{#1}%
\providecommand \bibfnamefont [1]{#1}%
\providecommand \citenamefont [1]{#1}%
\providecommand \href@noop [0]{\@secondoftwo}%
\providecommand \href [0]{\begingroup \@sanitize@url \@href}%
\providecommand \@href[1]{\@@startlink{#1}\@@href}%
\providecommand \@@href[1]{\endgroup#1\@@endlink}%
\providecommand \@sanitize@url [0]{\catcode `\\12\catcode `\$12\catcode
  `\&12\catcode `\#12\catcode `\^12\catcode `\_12\catcode `\%12\relax}%
\providecommand \@@startlink[1]{}%
\providecommand \@@endlink[0]{}%
\providecommand \url  [0]{\begingroup\@sanitize@url \@url }%
\providecommand \@url [1]{\endgroup\@href {#1}{\urlprefix }}%
\providecommand \urlprefix  [0]{URL }%
\providecommand \Eprint [0]{\href }%
\providecommand \doibase [0]{https://doi.org/}%
\providecommand \selectlanguage [0]{\@gobble}%
\providecommand \bibinfo  [0]{\@secondoftwo}%
\providecommand \bibfield  [0]{\@secondoftwo}%
\providecommand \translation [1]{[#1]}%
\providecommand \BibitemOpen [0]{}%
\providecommand \bibitemStop [0]{}%
\providecommand \bibitemNoStop [0]{.\EOS\space}%
\providecommand \EOS [0]{\spacefactor3000\relax}%
\providecommand \BibitemShut  [1]{\csname bibitem#1\endcsname}%
\let\auto@bib@innerbib\@empty
\bibitem [{\citenamefont {Lines}\ and\ \citenamefont
  {Glass}(2001)}]{LinesGlass}%
  \BibitemOpen
  \bibfield  {author} {\bibinfo {author} {\bibfnamefont {M.~E.}\ \bibnamefont
  {Lines}}\ and\ \bibinfo {author} {\bibfnamefont {A.~M.}\ \bibnamefont
  {Glass}},\ }\href@noop {} {\emph {\bibinfo {title} {Principles and
  applications of ferroelectrics and related materials}}}\ (\bibinfo
  {publisher} {Oxford university press},\ \bibinfo {year} {2001})\BibitemShut
  {NoStop}%
\bibitem [{\citenamefont {Jia}\ \emph {et~al.}(2006)\citenamefont {Jia},
  \citenamefont {Nagarajan}, \citenamefont {He}, \citenamefont {Houben},
  \citenamefont {Zhao}, \citenamefont {Ramesh}, \citenamefont {Urban},\ and\
  \citenamefont {Waser}}]{Jia2006}%
  \BibitemOpen
  \bibfield  {author} {\bibinfo {author} {\bibfnamefont {C.-L.}\ \bibnamefont
  {Jia}}, \bibinfo {author} {\bibfnamefont {V.}~\bibnamefont {Nagarajan}},
  \bibinfo {author} {\bibfnamefont {J.-Q.}\ \bibnamefont {He}}, \bibinfo
  {author} {\bibfnamefont {L.}~\bibnamefont {Houben}}, \bibinfo {author}
  {\bibfnamefont {T.}~\bibnamefont {Zhao}}, \bibinfo {author} {\bibfnamefont
  {R.}~\bibnamefont {Ramesh}}, \bibinfo {author} {\bibfnamefont
  {K.}~\bibnamefont {Urban}},\ and\ \bibinfo {author} {\bibfnamefont
  {R.}~\bibnamefont {Waser}},\ }\bibfield  {title} {\bibinfo {title} {Unit-cell
  scale mapping of ferroelectricity and tetragonality in epitaxial ultrathin
  ferroelectric films},\ }\href {https://doi.org/10.1038/nmat1808} {\bibfield
  {journal} {\bibinfo  {journal} {Nature Materials}\ }\textbf {\bibinfo
  {volume} {6}},\ \bibinfo {pages} {64–69} (\bibinfo {year}
  {2006})}\BibitemShut {NoStop}%
\bibitem [{\citenamefont {Fong}\ \emph {et~al.}(2004)\citenamefont {Fong},
  \citenamefont {Stephenson}, \citenamefont {Streiffer}, \citenamefont
  {Eastman}, \citenamefont {Auciello}, \citenamefont {Fuoss},\ and\
  \citenamefont {Thompson}}]{Fong2004}%
  \BibitemOpen
  \bibfield  {author} {\bibinfo {author} {\bibfnamefont {D.~D.}\ \bibnamefont
  {Fong}}, \bibinfo {author} {\bibfnamefont {G.~B.}\ \bibnamefont
  {Stephenson}}, \bibinfo {author} {\bibfnamefont {S.~K.}\ \bibnamefont
  {Streiffer}}, \bibinfo {author} {\bibfnamefont {J.~A.}\ \bibnamefont
  {Eastman}}, \bibinfo {author} {\bibfnamefont {O.}~\bibnamefont {Auciello}},
  \bibinfo {author} {\bibfnamefont {P.~H.}\ \bibnamefont {Fuoss}},\ and\
  \bibinfo {author} {\bibfnamefont {C.}~\bibnamefont {Thompson}},\ }\bibfield
  {title} {\bibinfo {title} {Ferroelectricity in ultrathin perovskite films},\
  }\href {https://doi.org/10.1126/science.1098252} {\bibfield  {journal}
  {\bibinfo  {journal} {Science}\ }\textbf {\bibinfo {volume} {304}},\ \bibinfo
  {pages} {1650–1653} (\bibinfo {year} {2004})}\BibitemShut {NoStop}%
\bibitem [{\citenamefont {Benedek}\ and\ \citenamefont
  {Fennie}(2011)}]{Benedek2011}%
  \BibitemOpen
  \bibfield  {author} {\bibinfo {author} {\bibfnamefont {N.~A.}\ \bibnamefont
  {Benedek}}\ and\ \bibinfo {author} {\bibfnamefont {C.~J.}\ \bibnamefont
  {Fennie}},\ }\bibfield  {title} {\bibinfo {title} {Hybrid improper
  ferroelectricity: a mechanism for controllable polarization-magnetization
  coupling},\ }\href@noop {} {\bibfield  {journal} {\bibinfo  {journal}
  {Physical Review Letters}\ }\textbf {\bibinfo {volume} {106}},\ \bibinfo
  {pages} {107204} (\bibinfo {year} {2011})}\BibitemShut {NoStop}%
\bibitem [{\citenamefont {Harris}(2011)}]{Harris2011}%
  \BibitemOpen
  \bibfield  {author} {\bibinfo {author} {\bibfnamefont {A.~B.}\ \bibnamefont
  {Harris}},\ }\bibfield  {title} {\bibinfo {title} {Symmetry analysis for the
  ruddlesden-popper systems \ce{Ca3Mn2O7} and \ce{Ca3Ti2O7}},\ }\bibfield
  {journal} {\bibinfo  {journal} {Physical Review B}\ }\textbf {\bibinfo
  {volume} {84}},\ \href {https://doi.org/10.1103/physrevb.84.064116}
  {10.1103/physrevb.84.064116} (\bibinfo {year} {2011})\BibitemShut {NoStop}%
\bibitem [{\citenamefont {Bousquet}\ \emph {et~al.}(2008)\citenamefont
  {Bousquet}, \citenamefont {Dawber}, \citenamefont {Stucki}, \citenamefont
  {Lichtensteiger}, \citenamefont {Hermet}, \citenamefont {Gariglio},
  \citenamefont {Triscone},\ and\ \citenamefont {Ghosez}}]{Bousquet2008}%
  \BibitemOpen
  \bibfield  {author} {\bibinfo {author} {\bibfnamefont {E.}~\bibnamefont
  {Bousquet}}, \bibinfo {author} {\bibfnamefont {M.}~\bibnamefont {Dawber}},
  \bibinfo {author} {\bibfnamefont {N.}~\bibnamefont {Stucki}}, \bibinfo
  {author} {\bibfnamefont {C.}~\bibnamefont {Lichtensteiger}}, \bibinfo
  {author} {\bibfnamefont {P.}~\bibnamefont {Hermet}}, \bibinfo {author}
  {\bibfnamefont {S.}~\bibnamefont {Gariglio}}, \bibinfo {author}
  {\bibfnamefont {J.-M.}\ \bibnamefont {Triscone}},\ and\ \bibinfo {author}
  {\bibfnamefont {P.}~\bibnamefont {Ghosez}},\ }\bibfield  {title} {\bibinfo
  {title} {Improper ferroelectricity in perovskite oxide artificial
  superlattices},\ }\href@noop {} {\bibfield  {journal} {\bibinfo  {journal}
  {Nature}\ }\textbf {\bibinfo {volume} {452}},\ \bibinfo {pages} {732}
  (\bibinfo {year} {2008})}\BibitemShut {NoStop}%
\bibitem [{\citenamefont {Garrity}\ \emph {et~al.}(2014)\citenamefont
  {Garrity}, \citenamefont {Rabe},\ and\ \citenamefont
  {Vanderbilt}}]{Garrity2014}%
  \BibitemOpen
  \bibfield  {author} {\bibinfo {author} {\bibfnamefont {K.~F.}\ \bibnamefont
  {Garrity}}, \bibinfo {author} {\bibfnamefont {K.~M.}\ \bibnamefont {Rabe}},\
  and\ \bibinfo {author} {\bibfnamefont {D.}~\bibnamefont {Vanderbilt}},\
  }\bibfield  {title} {\bibinfo {title} {Hyperferroelectrics: Proper
  ferroelectrics with persistent polarization},\ }\href
  {https://doi.org/10.1103/physrevlett.112.127601} {\bibfield  {journal}
  {\bibinfo  {journal} {Physical Review Letters}\ }\textbf {\bibinfo {volume}
  {112}},\ \bibinfo {pages} {127601} (\bibinfo {year} {2014})}\BibitemShut
  {NoStop}%
\bibitem [{\citenamefont {Martin}\ and\ \citenamefont
  {Rappe}(2016)}]{Martin2016}%
  \BibitemOpen
  \bibfield  {author} {\bibinfo {author} {\bibfnamefont {L.~W.}\ \bibnamefont
  {Martin}}\ and\ \bibinfo {author} {\bibfnamefont {A.~M.}\ \bibnamefont
  {Rappe}},\ }\bibfield  {title} {\bibinfo {title} {Thin-film ferroelectric
  materials and their applications},\ }\bibfield  {journal} {\bibinfo
  {journal} {Nature Reviews Materials}\ }\textbf {\bibinfo {volume} {2}},\
  \href {https://doi.org/10.1038/natrevmats.2016.87}
  {10.1038/natrevmats.2016.87} (\bibinfo {year} {2016})\BibitemShut {NoStop}%
\bibitem [{\citenamefont {Cheema}\ \emph {et~al.}(2020)\citenamefont {Cheema},
  \citenamefont {Kwon}, \citenamefont {Shanker}, \citenamefont {dos Reis},
  \citenamefont {Hsu}, \citenamefont {Xiao}, \citenamefont {Zhang},
  \citenamefont {Wagner}, \citenamefont {Datar}, \citenamefont {McCarter},
  \citenamefont {Serrao}, \citenamefont {Yadav}, \citenamefont {Karbasian},
  \citenamefont {Hsu}, \citenamefont {Tan}, \citenamefont {Wang}, \citenamefont
  {Thakare}, \citenamefont {Zhang}, \citenamefont {Mehta}, \citenamefont
  {Karapetrova}, \citenamefont {Chopdekar}, \citenamefont {Shafer},
  \citenamefont {Arenholz}, \citenamefont {Hu}, \citenamefont {Proksch},
  \citenamefont {Ramesh}, \citenamefont {Ciston},\ and\ \citenamefont
  {Salahuddin}}]{Cheema2020}%
  \BibitemOpen
  \bibfield  {author} {\bibinfo {author} {\bibfnamefont {S.~S.}\ \bibnamefont
  {Cheema}}, \bibinfo {author} {\bibfnamefont {D.}~\bibnamefont {Kwon}},
  \bibinfo {author} {\bibfnamefont {N.}~\bibnamefont {Shanker}}, \bibinfo
  {author} {\bibfnamefont {R.}~\bibnamefont {dos Reis}}, \bibinfo {author}
  {\bibfnamefont {S.-L.}\ \bibnamefont {Hsu}}, \bibinfo {author} {\bibfnamefont
  {J.}~\bibnamefont {Xiao}}, \bibinfo {author} {\bibfnamefont {H.}~\bibnamefont
  {Zhang}}, \bibinfo {author} {\bibfnamefont {R.}~\bibnamefont {Wagner}},
  \bibinfo {author} {\bibfnamefont {A.}~\bibnamefont {Datar}}, \bibinfo
  {author} {\bibfnamefont {M.~R.}\ \bibnamefont {McCarter}}, \bibinfo {author}
  {\bibfnamefont {C.~R.}\ \bibnamefont {Serrao}}, \bibinfo {author}
  {\bibfnamefont {A.~K.}\ \bibnamefont {Yadav}}, \bibinfo {author}
  {\bibfnamefont {G.}~\bibnamefont {Karbasian}}, \bibinfo {author}
  {\bibfnamefont {C.-H.}\ \bibnamefont {Hsu}}, \bibinfo {author} {\bibfnamefont
  {A.~J.}\ \bibnamefont {Tan}}, \bibinfo {author} {\bibfnamefont {L.-C.}\
  \bibnamefont {Wang}}, \bibinfo {author} {\bibfnamefont {V.}~\bibnamefont
  {Thakare}}, \bibinfo {author} {\bibfnamefont {X.}~\bibnamefont {Zhang}},
  \bibinfo {author} {\bibfnamefont {A.}~\bibnamefont {Mehta}}, \bibinfo
  {author} {\bibfnamefont {E.}~\bibnamefont {Karapetrova}}, \bibinfo {author}
  {\bibfnamefont {R.~V.}\ \bibnamefont {Chopdekar}}, \bibinfo {author}
  {\bibfnamefont {P.}~\bibnamefont {Shafer}}, \bibinfo {author} {\bibfnamefont
  {E.}~\bibnamefont {Arenholz}}, \bibinfo {author} {\bibfnamefont
  {C.}~\bibnamefont {Hu}}, \bibinfo {author} {\bibfnamefont {R.}~\bibnamefont
  {Proksch}}, \bibinfo {author} {\bibfnamefont {R.}~\bibnamefont {Ramesh}},
  \bibinfo {author} {\bibfnamefont {J.}~\bibnamefont {Ciston}},\ and\ \bibinfo
  {author} {\bibfnamefont {S.}~\bibnamefont {Salahuddin}},\ }\bibfield  {title}
  {\bibinfo {title} {Enhanced ferroelectricity in ultrathin films grown
  directly on silicon},\ }\href {https://doi.org/10.1038/s41586-020-2208-x}
  {\bibfield  {journal} {\bibinfo  {journal} {Nature}\ }\textbf {\bibinfo
  {volume} {580}},\ \bibinfo {pages} {478–482} (\bibinfo {year}
  {2020})}\BibitemShut {NoStop}%
\bibitem [{\citenamefont {Spreitzer}\ \emph {et~al.}(2021)\citenamefont
  {Spreitzer}, \citenamefont {Klement}, \citenamefont {Parkelj~Potočnik},
  \citenamefont {Trstenjak}, \citenamefont {Jovanović}, \citenamefont
  {Nguyen}, \citenamefont {Yuan}, \citenamefont {ten Elshof}, \citenamefont
  {Houwman}, \citenamefont {Koster},\ and\ \citenamefont
  {et~al.}}]{Spreitzer2021}%
  \BibitemOpen
  \bibfield  {author} {\bibinfo {author} {\bibfnamefont {M.}~\bibnamefont
  {Spreitzer}}, \bibinfo {author} {\bibfnamefont {D.}~\bibnamefont {Klement}},
  \bibinfo {author} {\bibfnamefont {T.}~\bibnamefont {Parkelj~Potočnik}},
  \bibinfo {author} {\bibfnamefont {U.}~\bibnamefont {Trstenjak}}, \bibinfo
  {author} {\bibfnamefont {Z.}~\bibnamefont {Jovanović}}, \bibinfo {author}
  {\bibfnamefont {M.~D.}\ \bibnamefont {Nguyen}}, \bibinfo {author}
  {\bibfnamefont {H.}~\bibnamefont {Yuan}}, \bibinfo {author} {\bibfnamefont
  {J.~E.}\ \bibnamefont {ten Elshof}}, \bibinfo {author} {\bibfnamefont
  {E.}~\bibnamefont {Houwman}}, \bibinfo {author} {\bibfnamefont
  {G.}~\bibnamefont {Koster}},\ and\ \bibinfo {author} {\bibnamefont
  {et~al.}},\ }\bibfield  {title} {\bibinfo {title} {Epitaxial ferroelectric
  oxides on silicon with perspectives for future device applications},\ }\href
  {https://doi.org/10.1063/5.0039161} {\bibfield  {journal} {\bibinfo
  {journal} {APL Materials}\ }\textbf {\bibinfo {volume} {9}},\ \bibinfo
  {pages} {040701} (\bibinfo {year} {2021})}\BibitemShut {NoStop}%
\bibitem [{\citenamefont {Ren}(2004)}]{Ren2004}%
  \BibitemOpen
  \bibfield  {author} {\bibinfo {author} {\bibfnamefont {X.}~\bibnamefont
  {Ren}},\ }\bibfield  {title} {\bibinfo {title} {Large electric-field-induced
  strain in ferroelectric crystals by point-defect-mediated reversible domain
  switching},\ }\href {https://doi.org/10.1038/nmat1051} {\bibfield  {journal}
  {\bibinfo  {journal} {Nature Materials}\ }\textbf {\bibinfo {volume} {3}},\
  \bibinfo {pages} {91–94} (\bibinfo {year} {2004})}\BibitemShut {NoStop}%
\bibitem [{\citenamefont {Li}\ and\ \citenamefont {Birol}(2021)}]{Li2021}%
  \BibitemOpen
  \bibfield  {author} {\bibinfo {author} {\bibfnamefont {S.}~\bibnamefont
  {Li}}\ and\ \bibinfo {author} {\bibfnamefont {T.}~\bibnamefont {Birol}},\
  }\bibfield  {title} {\bibinfo {title} {Free-carrier-induced ferroelectricity
  in layered perovskites},\ }\href
  {https://doi.org/10.1103/physrevlett.127.087601} {\bibfield  {journal}
  {\bibinfo  {journal} {Physical Review Letters}\ }\textbf {\bibinfo {volume}
  {127}},\ \bibinfo {pages} {087601} (\bibinfo {year} {2021})}\BibitemShut
  {NoStop}%
\bibitem [{\citenamefont {Kolodiazhnyi}\ \emph {et~al.}(2010)\citenamefont
  {Kolodiazhnyi}, \citenamefont {Tachibana}, \citenamefont {Kawaji},
  \citenamefont {Hwang},\ and\ \citenamefont
  {Takayama-Muromachi}}]{Kolodiazhnyi2010}%
  \BibitemOpen
  \bibfield  {author} {\bibinfo {author} {\bibfnamefont {T.}~\bibnamefont
  {Kolodiazhnyi}}, \bibinfo {author} {\bibfnamefont {M.}~\bibnamefont
  {Tachibana}}, \bibinfo {author} {\bibfnamefont {H.}~\bibnamefont {Kawaji}},
  \bibinfo {author} {\bibfnamefont {J.}~\bibnamefont {Hwang}},\ and\ \bibinfo
  {author} {\bibfnamefont {E.}~\bibnamefont {Takayama-Muromachi}},\ }\bibfield
  {title} {\bibinfo {title} {Persistence of ferroelectricity in
  \ce{BaTiO3}through the insulator-metal transition},\ }\href
  {https://doi.org/10.1103/physrevlett.104.147602} {\bibfield  {journal}
  {\bibinfo  {journal} {Physical Review Letters}\ }\textbf {\bibinfo {volume}
  {104}},\ \bibinfo {pages} {1470602} (\bibinfo {year} {2010})}\BibitemShut
  {NoStop}%
\bibitem [{\citenamefont {M{\"u}ller}\ \emph {et~al.}(2011)\citenamefont
  {M{\"u}ller}, \citenamefont {Schr{\"o}der}, \citenamefont {B{\"o}scke},
  \citenamefont {M{\"u}ller}, \citenamefont {B{\"o}ttger}, \citenamefont
  {Wilde}, \citenamefont {Sundqvist}, \citenamefont {Lemberger}, \citenamefont
  {K{\"u}cher}, \citenamefont {Mikolajick} \emph {et~al.}}]{Muller2011}%
  \BibitemOpen
  \bibfield  {author} {\bibinfo {author} {\bibfnamefont {J.}~\bibnamefont
  {M{\"u}ller}}, \bibinfo {author} {\bibfnamefont {U.}~\bibnamefont
  {Schr{\"o}der}}, \bibinfo {author} {\bibfnamefont {T.}~\bibnamefont
  {B{\"o}scke}}, \bibinfo {author} {\bibfnamefont {I.}~\bibnamefont
  {M{\"u}ller}}, \bibinfo {author} {\bibfnamefont {U.}~\bibnamefont
  {B{\"o}ttger}}, \bibinfo {author} {\bibfnamefont {L.}~\bibnamefont {Wilde}},
  \bibinfo {author} {\bibfnamefont {J.}~\bibnamefont {Sundqvist}}, \bibinfo
  {author} {\bibfnamefont {M.}~\bibnamefont {Lemberger}}, \bibinfo {author}
  {\bibfnamefont {P.}~\bibnamefont {K{\"u}cher}}, \bibinfo {author}
  {\bibfnamefont {T.}~\bibnamefont {Mikolajick}}, \emph {et~al.},\ }\bibfield
  {title} {\bibinfo {title} {Ferroelectricity in yttrium-doped hafnium oxide},\
  }\href {https://doi.org/10.1063/1.3667205} {\bibfield  {journal} {\bibinfo
  {journal} {Journal of Applied Physics}\ }\textbf {\bibinfo {volume} {110}},\
  \bibinfo {pages} {114113} (\bibinfo {year} {2011})}\BibitemShut {NoStop}%
\bibitem [{\citenamefont {Mueller}\ \emph {et~al.}(2012)\citenamefont
  {Mueller}, \citenamefont {Mueller}, \citenamefont {Singh}, \citenamefont
  {Riedel}, \citenamefont {Sundqvist}, \citenamefont {Schroeder},\ and\
  \citenamefont {Mikolajick}}]{Mueller2012}%
  \BibitemOpen
  \bibfield  {author} {\bibinfo {author} {\bibfnamefont {S.}~\bibnamefont
  {Mueller}}, \bibinfo {author} {\bibfnamefont {J.}~\bibnamefont {Mueller}},
  \bibinfo {author} {\bibfnamefont {A.}~\bibnamefont {Singh}}, \bibinfo
  {author} {\bibfnamefont {S.}~\bibnamefont {Riedel}}, \bibinfo {author}
  {\bibfnamefont {J.}~\bibnamefont {Sundqvist}}, \bibinfo {author}
  {\bibfnamefont {U.}~\bibnamefont {Schroeder}},\ and\ \bibinfo {author}
  {\bibfnamefont {T.}~\bibnamefont {Mikolajick}},\ }\bibfield  {title}
  {\bibinfo {title} {Incipient ferroelectricity in al-doped hfo2 thin films},\
  }\href {https://doi.org/10.1002/adfm.201103119} {\bibfield  {journal}
  {\bibinfo  {journal} {Advanced Functional Materials}\ }\textbf {\bibinfo
  {volume} {22}},\ \bibinfo {pages} {2412–2417} (\bibinfo {year}
  {2012})}\BibitemShut {NoStop}%
\bibitem [{\citenamefont {Hoffmann}\ \emph {et~al.}(2015)\citenamefont
  {Hoffmann}, \citenamefont {Schroeder}, \citenamefont {Schenk}, \citenamefont
  {Shimizu}, \citenamefont {Funakubo}, \citenamefont {Sakata}, \citenamefont
  {Pohl}, \citenamefont {Drescher}, \citenamefont {Adelmann}, \citenamefont
  {Materlik},\ and\ \citenamefont {et~al.}}]{Hoffmann2015}%
  \BibitemOpen
  \bibfield  {author} {\bibinfo {author} {\bibfnamefont {M.}~\bibnamefont
  {Hoffmann}}, \bibinfo {author} {\bibfnamefont {U.}~\bibnamefont {Schroeder}},
  \bibinfo {author} {\bibfnamefont {T.}~\bibnamefont {Schenk}}, \bibinfo
  {author} {\bibfnamefont {T.}~\bibnamefont {Shimizu}}, \bibinfo {author}
  {\bibfnamefont {H.}~\bibnamefont {Funakubo}}, \bibinfo {author}
  {\bibfnamefont {O.}~\bibnamefont {Sakata}}, \bibinfo {author} {\bibfnamefont
  {D.}~\bibnamefont {Pohl}}, \bibinfo {author} {\bibfnamefont {M.}~\bibnamefont
  {Drescher}}, \bibinfo {author} {\bibfnamefont {C.}~\bibnamefont {Adelmann}},
  \bibinfo {author} {\bibfnamefont {R.}~\bibnamefont {Materlik}},\ and\
  \bibinfo {author} {\bibnamefont {et~al.}},\ }\bibfield  {title} {\bibinfo
  {title} {Stabilizing the ferroelectric phase in doped hafnium oxide},\ }\href
  {https://doi.org/10.1063/1.4927805} {\bibfield  {journal} {\bibinfo
  {journal} {Journal of Applied Physics}\ }\textbf {\bibinfo {volume} {118}},\
  \bibinfo {pages} {072006} (\bibinfo {year} {2015})}\BibitemShut {NoStop}%
\bibitem [{\citenamefont {Ricca}\ \emph {et~al.}(2021)\citenamefont {Ricca},
  \citenamefont {Berkowitz},\ and\ \citenamefont {Aschauer}}]{Ricca2021}%
  \BibitemOpen
  \bibfield  {author} {\bibinfo {author} {\bibfnamefont {C.}~\bibnamefont
  {Ricca}}, \bibinfo {author} {\bibfnamefont {D.}~\bibnamefont {Berkowitz}},\
  and\ \bibinfo {author} {\bibfnamefont {U.}~\bibnamefont {Aschauer}},\
  }\bibfield  {title} {\bibinfo {title} {Ferroelectricity promoted by
  cation/anion divacancies in \ce{SrMnO3}},\ }\href
  {https://doi.org/10.1039/d1tc02317a} {\bibfield  {journal} {\bibinfo
  {journal} {Journal of Materials Chemistry C}\ ,\ \bibinfo {pages} {13321}}
  (\bibinfo {year} {2021})}\BibitemShut {NoStop}%
\bibitem [{\citenamefont {Hu}\ \emph {et~al.}(2013)\citenamefont {Hu},
  \citenamefont {Liu}, \citenamefont {Withers}, \citenamefont {Frankcombe},
  \citenamefont {Nor{\'{e}}n}, \citenamefont {Snashall}, \citenamefont
  {Kitchin}, \citenamefont {Smith}, \citenamefont {Gong}, \citenamefont {Chen},
  \citenamefont {Schiemer}, \citenamefont {Brink},\ and\ \citenamefont
  {Wong-Leung}}]{Hu2013}%
  \BibitemOpen
  \bibfield  {author} {\bibinfo {author} {\bibfnamefont {W.}~\bibnamefont
  {Hu}}, \bibinfo {author} {\bibfnamefont {Y.}~\bibnamefont {Liu}}, \bibinfo
  {author} {\bibfnamefont {R.~L.}\ \bibnamefont {Withers}}, \bibinfo {author}
  {\bibfnamefont {T.~J.}\ \bibnamefont {Frankcombe}}, \bibinfo {author}
  {\bibfnamefont {L.}~\bibnamefont {Nor{\'{e}}n}}, \bibinfo {author}
  {\bibfnamefont {A.}~\bibnamefont {Snashall}}, \bibinfo {author}
  {\bibfnamefont {M.}~\bibnamefont {Kitchin}}, \bibinfo {author} {\bibfnamefont
  {P.}~\bibnamefont {Smith}}, \bibinfo {author} {\bibfnamefont
  {B.}~\bibnamefont {Gong}}, \bibinfo {author} {\bibfnamefont {H.}~\bibnamefont
  {Chen}}, \bibinfo {author} {\bibfnamefont {J.}~\bibnamefont {Schiemer}},
  \bibinfo {author} {\bibfnamefont {F.}~\bibnamefont {Brink}},\ and\ \bibinfo
  {author} {\bibfnamefont {J.}~\bibnamefont {Wong-Leung}},\ }\bibfield  {title}
  {\bibinfo {title} {Electron-pinned defect-dipoles for high-performance
  colossal permittivity materials},\ }\href {https://doi.org/10.1038/nmat3691}
  {\bibfield  {journal} {\bibinfo  {journal} {Nature Materials}\ }\textbf
  {\bibinfo {volume} {12}},\ \bibinfo {pages} {821} (\bibinfo {year}
  {2013})}\BibitemShut {NoStop}%
\bibitem [{\citenamefont {Hu}\ \emph {et~al.}(2015)\citenamefont {Hu},
  \citenamefont {Lau}, \citenamefont {Liu}, \citenamefont {Withers},
  \citenamefont {Chen}, \citenamefont {Fu}, \citenamefont {Gong},\ and\
  \citenamefont {Hutchison}}]{Hu2015}%
  \BibitemOpen
  \bibfield  {author} {\bibinfo {author} {\bibfnamefont {W.}~\bibnamefont
  {Hu}}, \bibinfo {author} {\bibfnamefont {K.}~\bibnamefont {Lau}}, \bibinfo
  {author} {\bibfnamefont {Y.}~\bibnamefont {Liu}}, \bibinfo {author}
  {\bibfnamefont {R.~L.}\ \bibnamefont {Withers}}, \bibinfo {author}
  {\bibfnamefont {H.}~\bibnamefont {Chen}}, \bibinfo {author} {\bibfnamefont
  {L.}~\bibnamefont {Fu}}, \bibinfo {author} {\bibfnamefont {B.}~\bibnamefont
  {Gong}},\ and\ \bibinfo {author} {\bibfnamefont {W.}~\bibnamefont
  {Hutchison}},\ }\bibfield  {title} {\bibinfo {title} {Colossal dielectric
  permittivity in (\ce{Nb}+\ce{Al}) codoped rutile \ce{TiO2} ceramics:
  Compositional gradient and local structure},\ }\href
  {https://doi.org/10.1021/acs.chemmater.5b01351} {\bibfield  {journal}
  {\bibinfo  {journal} {Chemistry of Materials}\ }\textbf {\bibinfo {volume}
  {27}},\ \bibinfo {pages} {4934–4942} (\bibinfo {year} {2015})}\BibitemShut
  {NoStop}%
\bibitem [{\citenamefont {Dong}\ \emph {et~al.}(2015)\citenamefont {Dong},
  \citenamefont {Hu}, \citenamefont {Berlie}, \citenamefont {Lau},
  \citenamefont {Chen}, \citenamefont {Withers},\ and\ \citenamefont
  {Liu}}]{Dong2015}%
  \BibitemOpen
  \bibfield  {author} {\bibinfo {author} {\bibfnamefont {W.}~\bibnamefont
  {Dong}}, \bibinfo {author} {\bibfnamefont {W.}~\bibnamefont {Hu}}, \bibinfo
  {author} {\bibfnamefont {A.}~\bibnamefont {Berlie}}, \bibinfo {author}
  {\bibfnamefont {K.}~\bibnamefont {Lau}}, \bibinfo {author} {\bibfnamefont
  {H.}~\bibnamefont {Chen}}, \bibinfo {author} {\bibfnamefont {R.~L.}\
  \bibnamefont {Withers}},\ and\ \bibinfo {author} {\bibfnamefont
  {Y.}~\bibnamefont {Liu}},\ }\bibfield  {title} {\bibinfo {title} {Colossal
  dielectric behavior of \ce{Ga} + \ce{Nb} co-doped rutile \ce{TiO2}},\ }\href
  {https://doi.org/10.1021/acsami.5b07467} {\bibfield  {journal} {\bibinfo
  {journal} {{ACS} Applied Materials {\&} Interfaces}\ }\textbf {\bibinfo
  {volume} {7}},\ \bibinfo {pages} {25321} (\bibinfo {year}
  {2015})}\BibitemShut {NoStop}%
\bibitem [{\citenamefont {Bérardan}\ \emph {et~al.}(2016)\citenamefont
  {Bérardan}, \citenamefont {Franger}, \citenamefont {Dragoe}, \citenamefont
  {Meena},\ and\ \citenamefont {Dragoe}}]{Berardan2016}%
  \BibitemOpen
  \bibfield  {author} {\bibinfo {author} {\bibfnamefont {D.}~\bibnamefont
  {Bérardan}}, \bibinfo {author} {\bibfnamefont {S.}~\bibnamefont {Franger}},
  \bibinfo {author} {\bibfnamefont {D.}~\bibnamefont {Dragoe}}, \bibinfo
  {author} {\bibfnamefont {A.~K.}\ \bibnamefont {Meena}},\ and\ \bibinfo
  {author} {\bibfnamefont {N.}~\bibnamefont {Dragoe}},\ }\bibfield  {title}
  {\bibinfo {title} {Colossal dielectric constant in high entropy oxides},\
  }\href {https://doi.org/10.1002/pssr.201600043} {\bibfield  {journal}
  {\bibinfo  {journal} {physica status solidi (RRL) - Rapid Research Letters}\
  }\textbf {\bibinfo {volume} {10}},\ \bibinfo {pages} {328–333} (\bibinfo
  {year} {2016})}\BibitemShut {NoStop}%
\bibitem [{\citenamefont {Mulder}\ \emph {et~al.}(2013)\citenamefont {Mulder},
  \citenamefont {Benedek}, \citenamefont {Rondinelli},\ and\ \citenamefont
  {Fennie}}]{Mulder2013}%
  \BibitemOpen
  \bibfield  {author} {\bibinfo {author} {\bibfnamefont {A.~T.}\ \bibnamefont
  {Mulder}}, \bibinfo {author} {\bibfnamefont {N.~A.}\ \bibnamefont {Benedek}},
  \bibinfo {author} {\bibfnamefont {J.~M.}\ \bibnamefont {Rondinelli}},\ and\
  \bibinfo {author} {\bibfnamefont {C.~J.}\ \bibnamefont {Fennie}},\ }\bibfield
   {title} {\bibinfo {title} {Turning \ce{ABO3} antiferroelectrics into
  ferroelectrics: Design rules for practical rotation-driven ferroelectricity
  in double perovskites and \ce{A3B2O7} ruddlesden-popper compounds},\ }\href
  {https://doi.org/10.1002/adfm.201300210} {\bibfield  {journal} {\bibinfo
  {journal} {Advanced Functional Materials}\ ,\ \bibinfo {pages} {4810}}
  (\bibinfo {year} {2013})}\BibitemShut {NoStop}%
\bibitem [{\citenamefont {Resta}(1992)}]{Resta1992}%
  \BibitemOpen
  \bibfield  {author} {\bibinfo {author} {\bibfnamefont {R.}~\bibnamefont
  {Resta}},\ }\bibfield  {title} {\bibinfo {title} {Theory of the electric
  polarization in crystals},\ }\href@noop {} {\bibfield  {journal} {\bibinfo
  {journal} {Ferroelectrics}\ }\textbf {\bibinfo {volume} {136}},\ \bibinfo
  {pages} {51} (\bibinfo {year} {1992})}\BibitemShut {NoStop}%
\bibitem [{\citenamefont {King-Smith}\ and\ \citenamefont
  {Vanderbilt}(1993)}]{King-Smith1993}%
  \BibitemOpen
  \bibfield  {author} {\bibinfo {author} {\bibfnamefont {R.~D.}\ \bibnamefont
  {King-Smith}}\ and\ \bibinfo {author} {\bibfnamefont {D.}~\bibnamefont
  {Vanderbilt}},\ }\bibfield  {title} {\bibinfo {title} {Theory of polarization
  of crystalline solids},\ }\href@noop {} {\bibfield  {journal} {\bibinfo
  {journal} {Physical Review B}\ }\textbf {\bibinfo {volume} {47}},\ \bibinfo
  {pages} {1651} (\bibinfo {year} {1993})}\BibitemShut {NoStop}%
\bibitem [{\citenamefont {Bl{\"o}chl}(1994)}]{Blochl1994}%
  \BibitemOpen
  \bibfield  {author} {\bibinfo {author} {\bibfnamefont {P.~E.}\ \bibnamefont
  {Bl{\"o}chl}},\ }\bibfield  {title} {\bibinfo {title} {Projector
  augmented-wave method},\ }\href@noop {} {\bibfield  {journal} {\bibinfo
  {journal} {Physical Review B}\ }\textbf {\bibinfo {volume} {50}},\ \bibinfo
  {pages} {17953} (\bibinfo {year} {1994})}\BibitemShut {NoStop}%
\bibitem [{\citenamefont {Kresse}\ and\ \citenamefont
  {Furthm{\"u}ller}(1996{\natexlab{a}})}]{Kresse1996_PRB15}%
  \BibitemOpen
  \bibfield  {author} {\bibinfo {author} {\bibfnamefont {G.}~\bibnamefont
  {Kresse}}\ and\ \bibinfo {author} {\bibfnamefont {J.}~\bibnamefont
  {Furthm{\"u}ller}},\ }\bibfield  {title} {\bibinfo {title} {Efficiency of
  ab-initio total energy calculations for metals and semiconductors using a
  plane-wave basis set},\ }\href@noop {} {\bibfield  {journal} {\bibinfo
  {journal} {Computational Materials Science}\ }\textbf {\bibinfo {volume}
  {6}},\ \bibinfo {pages} {15} (\bibinfo {year}
  {1996}{\natexlab{a}})}\BibitemShut {NoStop}%
\bibitem [{\citenamefont {Kresse}\ and\ \citenamefont
  {Furthm{\"u}ller}(1996{\natexlab{b}})}]{Kresse1996_PRB11169}%
  \BibitemOpen
  \bibfield  {author} {\bibinfo {author} {\bibfnamefont {G.}~\bibnamefont
  {Kresse}}\ and\ \bibinfo {author} {\bibfnamefont {J.}~\bibnamefont
  {Furthm{\"u}ller}},\ }\bibfield  {title} {\bibinfo {title} {Efficient
  iterative schemes for ab initio total-energy calculations using a plane-wave
  basis set},\ }\href@noop {} {\bibfield  {journal} {\bibinfo  {journal}
  {Physical Review B}\ }\textbf {\bibinfo {volume} {54}},\ \bibinfo {pages}
  {11169} (\bibinfo {year} {1996}{\natexlab{b}})}\BibitemShut {NoStop}%
\bibitem [{\citenamefont {Heyd}\ \emph {et~al.}(2003)\citenamefont {Heyd},
  \citenamefont {Scuseria},\ and\ \citenamefont {Ernzerhof}}]{Heyd2003}%
  \BibitemOpen
  \bibfield  {author} {\bibinfo {author} {\bibfnamefont {J.}~\bibnamefont
  {Heyd}}, \bibinfo {author} {\bibfnamefont {G.~E.}\ \bibnamefont {Scuseria}},\
  and\ \bibinfo {author} {\bibfnamefont {M.}~\bibnamefont {Ernzerhof}},\
  }\bibfield  {title} {\bibinfo {title} {Hybrid functionals based on a screened
  coulomb potential},\ }\href@noop {} {\bibfield  {journal} {\bibinfo
  {journal} {The Journal of Chemical Physics}\ }\textbf {\bibinfo {volume}
  {118}},\ \bibinfo {pages} {8207} (\bibinfo {year} {2003})}\BibitemShut
  {NoStop}%
\bibitem [{Note1()}]{Note1}%
  \BibitemOpen
  \bibinfo {note} {The calculations of the defect formation energies, chemical
  potential phase diagrams and electrostatic convergence checks were done using
  the {\protect \it pydefect} package \cite {Kumagai2021}. The nudged elastic
  band calculations were done with {\protect \it VASP} modified with {\protect
  \it VTST} \cite {Sheppard2008}. The Hubbard $U$ parameter was applied through
  the rotational invariant form introduced by Dudarev et al.\cite
  {Dudarev1998}}\BibitemShut {NoStop}%
\bibitem [{\citenamefont {Momma}\ and\ \citenamefont
  {Izumi}(2011)}]{Momma2011}%
  \BibitemOpen
  \bibfield  {author} {\bibinfo {author} {\bibfnamefont {K.}~\bibnamefont
  {Momma}}\ and\ \bibinfo {author} {\bibfnamefont {F.}~\bibnamefont {Izumi}},\
  }\bibfield  {title} {\bibinfo {title} {Vesta 3 for three-dimensional
  visualization of crystal, volumetric and morphology data},\ }\href@noop {}
  {\bibfield  {journal} {\bibinfo  {journal} {Journal of Applied
  Crystallography}\ }\textbf {\bibinfo {volume} {44}},\ \bibinfo {pages} {1272}
  (\bibinfo {year} {2011})}\BibitemShut {NoStop}%
\bibitem [{\citenamefont {Reyes-Lillo}\ \emph {et~al.}(2016)\citenamefont
  {Reyes-Lillo}, \citenamefont {Rangel}, \citenamefont {Bruneval},\ and\
  \citenamefont {Neaton}}]{ReyesLillo2016}%
  \BibitemOpen
  \bibfield  {author} {\bibinfo {author} {\bibfnamefont {S.~E.}\ \bibnamefont
  {Reyes-Lillo}}, \bibinfo {author} {\bibfnamefont {T.}~\bibnamefont {Rangel}},
  \bibinfo {author} {\bibfnamefont {F.}~\bibnamefont {Bruneval}},\ and\
  \bibinfo {author} {\bibfnamefont {J.~B.}\ \bibnamefont {Neaton}},\ }\bibfield
   {title} {\bibinfo {title} {Effects of quantum confinement on excited state
  properties ofsrtio3fromab initiomany-body perturbation theory},\ }\href
  {https://doi.org/10.1103/physrevb.94.041107} {\bibfield  {journal} {\bibinfo
  {journal} {Physical Review B}\ }\textbf {\bibinfo {volume} {94}},\ \bibinfo
  {pages} {041107} (\bibinfo {year} {2016})}\BibitemShut {NoStop}%
\bibitem [{\citenamefont {Li}\ \emph {et~al.}(2019)\citenamefont {Li},
  \citenamefont {Niu}, \citenamefont {Zhao}, \citenamefont {Haiges},
  \citenamefont {Zhang}, \citenamefont {Ravichandran},\ and\ \citenamefont
  {Janotti}}]{Li2019}%
  \BibitemOpen
  \bibfield  {author} {\bibinfo {author} {\bibfnamefont {W.}~\bibnamefont
  {Li}}, \bibinfo {author} {\bibfnamefont {S.}~\bibnamefont {Niu}}, \bibinfo
  {author} {\bibfnamefont {B.}~\bibnamefont {Zhao}}, \bibinfo {author}
  {\bibfnamefont {R.}~\bibnamefont {Haiges}}, \bibinfo {author} {\bibfnamefont
  {Z.}~\bibnamefont {Zhang}}, \bibinfo {author} {\bibfnamefont
  {J.}~\bibnamefont {Ravichandran}},\ and\ \bibinfo {author} {\bibfnamefont
  {A.}~\bibnamefont {Janotti}},\ }\bibfield  {title} {\bibinfo {title} {Band
  gap evolution in ruddlesden-popper phases},\ }\href
  {https://doi.org/10.1103/physrevmaterials.3.101601} {\bibfield  {journal}
  {\bibinfo  {journal} {Physical Review Materials}\ }\textbf {\bibinfo {volume}
  {3}},\ \bibinfo {pages} {101601} (\bibinfo {year} {2019})}\BibitemShut
  {NoStop}%
\bibitem [{\citenamefont {Schultz}(1959)}]{Schultz1959}%
  \BibitemOpen
  \bibfield  {author} {\bibinfo {author} {\bibfnamefont {T.~D.}\ \bibnamefont
  {Schultz}},\ }\bibfield  {title} {\bibinfo {title} {Slow electrons in polar
  crystals: Self-energy, mass, and mobility},\ }\href
  {https://doi.org/10.1103/physrev.116.526} {\bibfield  {journal} {\bibinfo
  {journal} {Physical Review}\ }\textbf {\bibinfo {volume} {116}},\ \bibinfo
  {pages} {526–543} (\bibinfo {year} {1959})}\BibitemShut {NoStop}%
\bibitem [{\citenamefont {Feynman}(1955)}]{Feynman1955}%
  \BibitemOpen
  \bibfield  {author} {\bibinfo {author} {\bibfnamefont {R.~P.}\ \bibnamefont
  {Feynman}},\ }\bibfield  {title} {\bibinfo {title} {Slow electrons in a polar
  crystal},\ }\href {https://doi.org/10.1103/physrev.97.660} {\bibfield
  {journal} {\bibinfo  {journal} {Physical Review}\ }\textbf {\bibinfo {volume}
  {97}},\ \bibinfo {pages} {660–665} (\bibinfo {year} {1955})}\BibitemShut
  {NoStop}%
\bibitem [{\citenamefont {Frost}(2017)}]{Frost2017}%
  \BibitemOpen
  \bibfield  {author} {\bibinfo {author} {\bibfnamefont {J.~M.}\ \bibnamefont
  {Frost}},\ }\bibfield  {title} {\bibinfo {title} {Calculating polaron
  mobility in halide perovskites},\ }\href
  {https://doi.org/10.1103/physrevb.96.195202} {\bibfield  {journal} {\bibinfo
  {journal} {Physical Review B}\ }\textbf {\bibinfo {volume} {96}},\ \bibinfo
  {pages} {195202} (\bibinfo {year} {2017})}\BibitemShut {NoStop}%
\bibitem [{\citenamefont {Togo}\ and\ \citenamefont {Tanaka}(2015)}]{Togo2015}%
  \BibitemOpen
  \bibfield  {author} {\bibinfo {author} {\bibfnamefont {A.}~\bibnamefont
  {Togo}}\ and\ \bibinfo {author} {\bibfnamefont {I.}~\bibnamefont {Tanaka}},\
  }\bibfield  {title} {\bibinfo {title} {First principles phonon calculations
  in materials science},\ }\href@noop {} {\bibfield  {journal} {\bibinfo
  {journal} {Scripta Materialia}\ }\textbf {\bibinfo {volume} {108}},\ \bibinfo
  {pages} {1} (\bibinfo {year} {2015})}\BibitemShut {NoStop}%
\bibitem [{\citenamefont {Skelton}\ \emph {et~al.}(2017)\citenamefont
  {Skelton}, \citenamefont {Burton}, \citenamefont {Jackson}, \citenamefont
  {Oba}, \citenamefont {Parker},\ and\ \citenamefont {Walsh}}]{Skelton2017}%
  \BibitemOpen
  \bibfield  {author} {\bibinfo {author} {\bibfnamefont {J.~M.}\ \bibnamefont
  {Skelton}}, \bibinfo {author} {\bibfnamefont {L.~A.}\ \bibnamefont {Burton}},
  \bibinfo {author} {\bibfnamefont {A.~J.}\ \bibnamefont {Jackson}}, \bibinfo
  {author} {\bibfnamefont {F.}~\bibnamefont {Oba}}, \bibinfo {author}
  {\bibfnamefont {S.~C.}\ \bibnamefont {Parker}},\ and\ \bibinfo {author}
  {\bibfnamefont {A.}~\bibnamefont {Walsh}},\ }\bibfield  {title} {\bibinfo
  {title} {Lattice dynamics of the tin sulphides \ce{SnS2}, \ce{SnS} and
  \ce{Sn2S3}: vibrational spectra and thermal transport},\ }\href@noop {}
  {\bibfield  {journal} {\bibinfo  {journal} {Physical Chemistry Chemical
  Physics}\ }\textbf {\bibinfo {volume} {19}},\ \bibinfo {pages} {12452}
  (\bibinfo {year} {2017})}\BibitemShut {NoStop}%
\bibitem [{\citenamefont {Hellwarth}\ and\ \citenamefont
  {Biaggio}(1999)}]{Hellwarth1999}%
  \BibitemOpen
  \bibfield  {author} {\bibinfo {author} {\bibfnamefont {R.~W.}\ \bibnamefont
  {Hellwarth}}\ and\ \bibinfo {author} {\bibfnamefont {I.}~\bibnamefont
  {Biaggio}},\ }\bibfield  {title} {\bibinfo {title} {Mobility of an electron
  in a multimode polar lattice},\ }\href
  {https://doi.org/10.1103/physrevb.60.299} {\bibfield  {journal} {\bibinfo
  {journal} {Physical Review B}\ }\textbf {\bibinfo {volume} {60}},\ \bibinfo
  {pages} {299–307} (\bibinfo {year} {1999})}\BibitemShut {NoStop}%
\bibitem [{\citenamefont {Sio}\ \emph {et~al.}(2019)\citenamefont {Sio},
  \citenamefont {Verdi}, \citenamefont {Poncé},\ and\ \citenamefont
  {Giustino}}]{Sio2019}%
  \BibitemOpen
  \bibfield  {author} {\bibinfo {author} {\bibfnamefont {W.~H.}\ \bibnamefont
  {Sio}}, \bibinfo {author} {\bibfnamefont {C.}~\bibnamefont {Verdi}}, \bibinfo
  {author} {\bibfnamefont {S.}~\bibnamefont {Poncé}},\ and\ \bibinfo {author}
  {\bibfnamefont {F.}~\bibnamefont {Giustino}},\ }\bibfield  {title} {\bibinfo
  {title} {Ab initiotheory of polarons: Formalism and applications},\ }\href
  {https://doi.org/10.1103/physrevb.99.235139} {\bibfield  {journal} {\bibinfo
  {journal} {Physical Review B}\ }\textbf {\bibinfo {volume} {99}},\ \bibinfo
  {pages} {235139} (\bibinfo {year} {2019})}\BibitemShut {NoStop}%
\bibitem [{\citenamefont {Devreese}\ and\ \citenamefont
  {Alexandrov}(2009)}]{Devreese2009}%
  \BibitemOpen
  \bibfield  {author} {\bibinfo {author} {\bibfnamefont {J.~T.}\ \bibnamefont
  {Devreese}}\ and\ \bibinfo {author} {\bibfnamefont {A.~S.}\ \bibnamefont
  {Alexandrov}},\ }\bibfield  {title} {\bibinfo {title} {Fröhlich polaron and
  bipolaron: recent developments},\ }\href
  {https://doi.org/10.1088/0034-4885/72/6/066501} {\bibfield  {journal}
  {\bibinfo  {journal} {Reports on Progress in Physics}\ }\textbf {\bibinfo
  {volume} {72}},\ \bibinfo {pages} {066501} (\bibinfo {year}
  {2009})}\BibitemShut {NoStop}%
\bibitem [{\citenamefont {Vanderbilt}\ and\ \citenamefont
  {King-Smith}(1993)}]{Vanderbilt1993}%
  \BibitemOpen
  \bibfield  {author} {\bibinfo {author} {\bibfnamefont {D.}~\bibnamefont
  {Vanderbilt}}\ and\ \bibinfo {author} {\bibfnamefont {R.~D.}\ \bibnamefont
  {King-Smith}},\ }\bibfield  {title} {\bibinfo {title} {Electric polarization
  as a bulk quantity and its relation to surface charge},\ }\href
  {https://doi.org/10.1103/physrevb.48.4442} {\bibfield  {journal} {\bibinfo
  {journal} {Physical Review B}\ }\textbf {\bibinfo {volume} {48}},\ \bibinfo
  {pages} {4442–4455} (\bibinfo {year} {1993})}\BibitemShut {NoStop}%
\bibitem [{\citenamefont {Marton}\ \emph {et~al.}(2010)\citenamefont {Marton},
  \citenamefont {Rychetsky},\ and\ \citenamefont {Hlinka}}]{Marton2010}%
  \BibitemOpen
  \bibfield  {author} {\bibinfo {author} {\bibfnamefont {P.}~\bibnamefont
  {Marton}}, \bibinfo {author} {\bibfnamefont {I.}~\bibnamefont {Rychetsky}},\
  and\ \bibinfo {author} {\bibfnamefont {J.}~\bibnamefont {Hlinka}},\
  }\bibfield  {title} {\bibinfo {title} {Domain walls of ferroelectric
  \ce{BaTiO3} within the ginzburg-landau-devonshire phenomenological model},\
  }\href {https://doi.org/10.1103/physrevb.81.144125} {\bibfield  {journal}
  {\bibinfo  {journal} {Physical Review B}\ }\textbf {\bibinfo {volume} {81}},\
  \bibinfo {pages} {144125} (\bibinfo {year} {2010})}\BibitemShut {NoStop}%
\bibitem [{\citenamefont {Grünebohm}\ \emph {et~al.}(2012)\citenamefont
  {Grünebohm}, \citenamefont {Gruner},\ and\ \citenamefont
  {Entel}}]{Grunebohm2012}%
  \BibitemOpen
  \bibfield  {author} {\bibinfo {author} {\bibfnamefont {A.}~\bibnamefont
  {Grünebohm}}, \bibinfo {author} {\bibfnamefont {M.~E.}\ \bibnamefont
  {Gruner}},\ and\ \bibinfo {author} {\bibfnamefont {P.}~\bibnamefont
  {Entel}},\ }\bibfield  {title} {\bibinfo {title} {Domain structure in the
  tetragonal phase of \ce{BaTiO3} – from bulk to nanoparticles},\ }\href
  {https://doi.org/10.1080/00150193.2012.671090} {\bibfield  {journal}
  {\bibinfo  {journal} {Ferroelectrics}\ }\textbf {\bibinfo {volume} {426}},\
  \bibinfo {pages} {21–30} (\bibinfo {year} {2012})}\BibitemShut {NoStop}%
\bibitem [{\citenamefont {Grünebohm}\ and\ \citenamefont
  {Marathe}(2020)}]{Grunebohm2020}%
  \BibitemOpen
  \bibfield  {author} {\bibinfo {author} {\bibfnamefont {A.}~\bibnamefont
  {Grünebohm}}\ and\ \bibinfo {author} {\bibfnamefont {M.}~\bibnamefont
  {Marathe}},\ }\bibfield  {title} {\bibinfo {title} {Impact of domains on the
  orthorhombic-tetragonal transition of \ce{BaTiO3} : An ab initio study},\
  }\href {https://doi.org/10.1103/physrevmaterials.4.114417} {\bibfield
  {journal} {\bibinfo  {journal} {Physical Review Materials}\ }\textbf
  {\bibinfo {volume} {4}},\ \bibinfo {pages} {114417} (\bibinfo {year}
  {2020})}\BibitemShut {NoStop}%
\bibitem [{\citenamefont {Torrance}\ \emph {et~al.}(1991)\citenamefont
  {Torrance}, \citenamefont {Lacorre}, \citenamefont {Asavaroengchai},\ and\
  \citenamefont {Metzger}}]{Torrance1991}%
  \BibitemOpen
  \bibfield  {author} {\bibinfo {author} {\bibfnamefont {J.~B.}\ \bibnamefont
  {Torrance}}, \bibinfo {author} {\bibfnamefont {P.}~\bibnamefont {Lacorre}},
  \bibinfo {author} {\bibfnamefont {C.}~\bibnamefont {Asavaroengchai}},\ and\
  \bibinfo {author} {\bibfnamefont {R.~M.}\ \bibnamefont {Metzger}},\
  }\bibfield  {title} {\bibinfo {title} {Why are some oxides metallic, while
  most are insulating?},\ }\href {https://doi.org/10.1016/0921-4534(91)90534-6}
  {\bibfield  {journal} {\bibinfo  {journal} {Physica C: Superconductivity}\
  }\textbf {\bibinfo {volume} {182}},\ \bibinfo {pages} {351–364} (\bibinfo
  {year} {1991})}\BibitemShut {NoStop}%
\bibitem [{\citenamefont {Imada}\ \emph {et~al.}(1998)\citenamefont {Imada},
  \citenamefont {Fujimori},\ and\ \citenamefont {Tokura}}]{Imada1998}%
  \BibitemOpen
  \bibfield  {author} {\bibinfo {author} {\bibfnamefont {M.}~\bibnamefont
  {Imada}}, \bibinfo {author} {\bibfnamefont {A.}~\bibnamefont {Fujimori}},\
  and\ \bibinfo {author} {\bibfnamefont {Y.}~\bibnamefont {Tokura}},\
  }\bibfield  {title} {\bibinfo {title} {Metal-insulator transitions},\ }\href
  {https://doi.org/10.1103/revmodphys.70.1039} {\bibfield  {journal} {\bibinfo
  {journal} {Reviews of Modern Physics}\ }\textbf {\bibinfo {volume} {70}},\
  \bibinfo {pages} {1039–1263} (\bibinfo {year} {1998})}\BibitemShut
  {NoStop}%
\bibitem [{\citenamefont {Aryasetiawan}\ \emph {et~al.}(2006)\citenamefont
  {Aryasetiawan}, \citenamefont {Karlsson}, \citenamefont {Jepsen},\ and\
  \citenamefont {Sch{\"o}nberger}}]{Aryasetiawan2006}%
  \BibitemOpen
  \bibfield  {author} {\bibinfo {author} {\bibfnamefont {F.}~\bibnamefont
  {Aryasetiawan}}, \bibinfo {author} {\bibfnamefont {K.}~\bibnamefont
  {Karlsson}}, \bibinfo {author} {\bibfnamefont {O.}~\bibnamefont {Jepsen}},\
  and\ \bibinfo {author} {\bibfnamefont {U.}~\bibnamefont {Sch{\"o}nberger}},\
  }\bibfield  {title} {\bibinfo {title} {Calculations of hubbard u from
  first-principles},\ }\href@noop {} {\bibfield  {journal} {\bibinfo  {journal}
  {Physical Review B}\ }\textbf {\bibinfo {volume} {74}},\ \bibinfo {pages}
  {125106} (\bibinfo {year} {2006})}\BibitemShut {NoStop}%
\bibitem [{\citenamefont {Wehling}\ \emph {et~al.}(2011)\citenamefont
  {Wehling}, \citenamefont {Şaşıoğlu}, \citenamefont {Friedrich},
  \citenamefont {Lichtenstein}, \citenamefont {Katsnelson},\ and\ \citenamefont
  {Blügel}}]{Wehling2011}%
  \BibitemOpen
  \bibfield  {author} {\bibinfo {author} {\bibfnamefont {T.~O.}\ \bibnamefont
  {Wehling}}, \bibinfo {author} {\bibfnamefont {E.}~\bibnamefont
  {Şaşıoğlu}}, \bibinfo {author} {\bibfnamefont {C.}~\bibnamefont
  {Friedrich}}, \bibinfo {author} {\bibfnamefont {A.~I.}\ \bibnamefont
  {Lichtenstein}}, \bibinfo {author} {\bibfnamefont {M.~I.}\ \bibnamefont
  {Katsnelson}},\ and\ \bibinfo {author} {\bibfnamefont {S.}~\bibnamefont
  {Blügel}},\ }\bibfield  {title} {\bibinfo {title} {Strength of effective
  coulomb interactions in graphene and graphite},\ }\href
  {https://doi.org/10.1103/physrevlett.106.236805} {\bibfield  {journal}
  {\bibinfo  {journal} {Physical Review Letters}\ }\textbf {\bibinfo {volume}
  {106}},\ \bibinfo {pages} {236805} (\bibinfo {year} {2011})}\BibitemShut
  {NoStop}%
\bibitem [{\citenamefont {Janotti}\ \emph {et~al.}(2014)\citenamefont
  {Janotti}, \citenamefont {Varley}, \citenamefont {Choi},\ and\ \citenamefont
  {Van~de Walle}}]{Janotti2014}%
  \BibitemOpen
  \bibfield  {author} {\bibinfo {author} {\bibfnamefont {A.}~\bibnamefont
  {Janotti}}, \bibinfo {author} {\bibfnamefont {J.~B.}\ \bibnamefont {Varley}},
  \bibinfo {author} {\bibfnamefont {M.}~\bibnamefont {Choi}},\ and\ \bibinfo
  {author} {\bibfnamefont {C.~G.}\ \bibnamefont {Van~de Walle}},\ }\bibfield
  {title} {\bibinfo {title} {Vacancies and small polarons in \ce{SrTiO3}},\
  }\href {https://doi.org/10.1103/physrevb.90.085202} {\bibfield  {journal}
  {\bibinfo  {journal} {Physical Review B}\ }\textbf {\bibinfo {volume} {90}},\
  \bibinfo {pages} {085202} (\bibinfo {year} {2014})}\BibitemShut {NoStop}%
\bibitem [{\citenamefont {Haldane}\ and\ \citenamefont
  {Anderson}(1976)}]{Haldane1976}%
  \BibitemOpen
  \bibfield  {author} {\bibinfo {author} {\bibfnamefont {F.~D.~M.}\
  \bibnamefont {Haldane}}\ and\ \bibinfo {author} {\bibfnamefont {P.~W.}\
  \bibnamefont {Anderson}},\ }\bibfield  {title} {\bibinfo {title} {Simple
  model of multiple charge states of transition-metal impurities in
  semiconductors},\ }\href {https://doi.org/10.1103/physrevb.13.2553}
  {\bibfield  {journal} {\bibinfo  {journal} {Physical Review B}\ }\textbf
  {\bibinfo {volume} {13}},\ \bibinfo {pages} {2553–2559} (\bibinfo {year}
  {1976})}\BibitemShut {NoStop}%
\bibitem [{\citenamefont {Moritomo}\ \emph {et~al.}(1995)\citenamefont
  {Moritomo}, \citenamefont {Tomioka}, \citenamefont {Asamitsu}, \citenamefont
  {Tokura},\ and\ \citenamefont {Matsui}}]{Moritomo1995}%
  \BibitemOpen
  \bibfield  {author} {\bibinfo {author} {\bibfnamefont {Y.}~\bibnamefont
  {Moritomo}}, \bibinfo {author} {\bibfnamefont {Y.}~\bibnamefont {Tomioka}},
  \bibinfo {author} {\bibfnamefont {A.}~\bibnamefont {Asamitsu}}, \bibinfo
  {author} {\bibfnamefont {Y.}~\bibnamefont {Tokura}},\ and\ \bibinfo {author}
  {\bibfnamefont {Y.}~\bibnamefont {Matsui}},\ }\bibfield  {title} {\bibinfo
  {title} {Magnetic and electronic properties in hole-doped manganese oxides
  with layered structures \ce{La_{1-x}Sr_{1+x}MnO4}},\ }\href
  {https://doi.org/10.1103/physrevb.51.3297} {\bibfield  {journal} {\bibinfo
  {journal} {Physical Review B}\ }\textbf {\bibinfo {volume} {51}},\ \bibinfo
  {pages} {3297–3300} (\bibinfo {year} {1995})}\BibitemShut {NoStop}%
\bibitem [{\citenamefont {Kumagai}\ \emph {et~al.}(2021)\citenamefont
  {Kumagai}, \citenamefont {Tsunoda}, \citenamefont {Takahashi},\ and\
  \citenamefont {Oba}}]{Kumagai2021}%
  \BibitemOpen
  \bibfield  {author} {\bibinfo {author} {\bibfnamefont {Y.}~\bibnamefont
  {Kumagai}}, \bibinfo {author} {\bibfnamefont {N.}~\bibnamefont {Tsunoda}},
  \bibinfo {author} {\bibfnamefont {A.}~\bibnamefont {Takahashi}},\ and\
  \bibinfo {author} {\bibfnamefont {F.}~\bibnamefont {Oba}},\ }\bibfield
  {title} {\bibinfo {title} {Insights into oxygen vacancies from
  high-throughput first-principles calculations},\ }\href
  {http://dx.doi.org/10.1103/PhysRevMaterials.5.123803} {\bibfield  {journal}
  {\bibinfo  {journal} {Physical Review Materials}\ }\textbf {\bibinfo {volume}
  {5}} (\bibinfo {year} {2021})}\BibitemShut {NoStop}%
\bibitem [{\citenamefont {Sheppard}\ \emph {et~al.}(2008)\citenamefont
  {Sheppard}, \citenamefont {Terrell},\ and\ \citenamefont
  {Henkelman}}]{Sheppard2008}%
  \BibitemOpen
  \bibfield  {author} {\bibinfo {author} {\bibfnamefont {D.}~\bibnamefont
  {Sheppard}}, \bibinfo {author} {\bibfnamefont {R.}~\bibnamefont {Terrell}},\
  and\ \bibinfo {author} {\bibfnamefont {G.}~\bibnamefont {Henkelman}},\
  }\bibfield  {title} {\bibinfo {title} {Optimization methods for finding
  minimum energy paths},\ }\href {https://doi.org/10.1063/1.2841941} {\bibfield
   {journal} {\bibinfo  {journal} {The Journal of Chemical Physics}\ }\textbf
  {\bibinfo {volume} {128}},\ \bibinfo {pages} {134106} (\bibinfo {year}
  {2008})}\BibitemShut {NoStop}%
\bibitem [{\citenamefont {Dudarev}\ \emph {et~al.}(1998)\citenamefont
  {Dudarev}, \citenamefont {Botton}, \citenamefont {Savrasov}, \citenamefont
  {Humphreys},\ and\ \citenamefont {Sutton}}]{Dudarev1998}%
  \BibitemOpen
  \bibfield  {author} {\bibinfo {author} {\bibfnamefont {S.~L.}\ \bibnamefont
  {Dudarev}}, \bibinfo {author} {\bibfnamefont {G.~A.}\ \bibnamefont {Botton}},
  \bibinfo {author} {\bibfnamefont {S.~Y.}\ \bibnamefont {Savrasov}}, \bibinfo
  {author} {\bibfnamefont {C.~J.}\ \bibnamefont {Humphreys}},\ and\ \bibinfo
  {author} {\bibfnamefont {A.}~\bibnamefont {Sutton}},\ }\bibfield  {title}
  {\bibinfo {title} {Electron-energy-loss spectra and the structural stability
  of nickel oxide: An lsda+ u study},\ }\href@noop {} {\bibfield  {journal}
  {\bibinfo  {journal} {Physical Review B}\ }\textbf {\bibinfo {volume} {57}},\
  \bibinfo {pages} {1505} (\bibinfo {year} {1998})}\BibitemShut {NoStop}%
\end{thebibliography}%

\customlabel{figSI:prim}{S1}
\customlabel{figSI:phonons}{S2}
\customlabel{figSI:CRPA}{S3}
\customlabel{figSI:polaron_PBEU}{S4}
\customlabel{figSI:PBEU40_pol}{S5}
\customlabel{figSI:PBEU50_pol}{S6}
\customlabel{figSI:double_defect_all}{S7}
\customlabel{figSI:double_defect_integrate}{S8}
\customlabel{figSI:Fsite}{S9}
\customlabel{figSI:chempotHSE06}{S10}
\customlabel{figSI:chempot}{S11}
\customlabel{figSI:HSE06_elec}{S12}
\customlabel{figSI:PBEU40_elec}{S13}
\customlabel{figSI:PBEU50_elec}{S14}
\customlabel{figSI:PBEU_size}{S15}
\customlabel{figSI:full_dos}{S16}
\customlabel{figSI:polaron_density}{S17}
\customlabel{tabSI:dielectric_tensor}{S1}
\customlabel{fig:importance}{S1}

\end{document}